\documentclass[12pt]{article}

\usepackage{amssymb, latexsym, amsmath}
\usepackage[bookmarks = true]{hyperref}
\usepackage[dvips]{graphics}
\makeindex

\begin{document}

\title{\textbf{Pseudoduality and Conserved Currents in Sigma Models}}

\author{\textbf{Mustafa Sarisaman}\footnote{msarisaman@physics.miami.edu}}

\date{}

\maketitle
\par
\begin{center}
\textit{Department of Physics\\ University of Miami\\ P.O. Box 248046\\
 Coral Gables, FL 33124 USA}
\end{center}

\

\begin{center}
Monday, April 23, 2007
\end{center}

\

\begin{abstract}
We discuss the current conservation laws in sigma models based on a
compact Lie groups of the same dimensionality and connected to each
other via pseudoduality transformations in two dimensions. We show
that pseudoduality transformations induce an infinite number of
nonlocal conserved currents on the pseudodual manifold.
\end{abstract}

\vfill

\section{Introduction}\label{sec:int}

\par In this article we investigate some properties of conserved currents on target space manifolds that are pseudodual to each
other, following a method discussed in \cite{Alvarez1} we find an
infinite number of conservation laws in the pseudodual manifold. We
work out the currents in case of pseudodualities \cite{Curtright1,
Alvarez2} between the sigma model on an abelian group and a strict
WZW sigma model \cite{Witten} on a compact Lie group \cite{Evans1,
Evans2} of the same dimensionality. We specialize to the case of the
abelian group $U(1)\times U(1)\times U(1)$, and of the Lie group
$SU(2)$.

The strict WZW model is the model with the Wess-Zumino term
normalized so that the canonical equations of motion are given by
$\partial_{-} (g^{-1}\partial_{+}g) = 0$, where $g$ is a function on
spacetime taking values in some compact Lie group $G$. Pseudoduality
is defined as the "on shell" duality transformation mapping the
solutions to the equations of motion for different sigma models.
This transformation preserves the stress energy tensor though it is
not a canonical transformation. We know \cite{Alvarez1, Ivanov} that
if we are given a sigma model on an abelian group, and a strict WZW
sigma model on a compact Lie group, there is a duality
transformation between these two manifolds that maps solutions of
the equations of motion of the first manifold into the solutions of
the equations of motion of the second manifold. Solutions of the
equations of motions allow us to construct holomorphic \cite{Evans1}
nonlocal conserved currents on these manifolds. Pseudoduality
relations provide a way to form pseudodual currents, and we show
that these currents are conserved.

Let $\Sigma$ be two dimensional Minkowski space, and $\sigma^{\pm} = \tau \pm \sigma$ be the standard lightcone
coordinates. Using maps $x : \Sigma \rightarrow M$ and $\tilde{x} : \Sigma \rightarrow \tilde{M}$ we may
 write pseudoduality relations as
$$\tilde{x}_{+}(\sigma) = +T (\sigma)x_{+}(\sigma) \eqno(1.1) \label{equa1}$$
$$\tilde{x}_{-}(\sigma) = -T (\sigma)x_{-}(\sigma) \eqno(1.2)$$
where $T(\sigma)$ belongs to $SO(n)$, and is a function of $\sigma$.

Let $M = G$ be a compact Lie group of dimension $n$ with an
Ad(G)-invariant metric, and $g : \Sigma \rightarrow G$. We define
the basic nonlocal conserved currents $J_{+}^{(L)} =
(g^{-1}\partial_{+}g)$ and $J_{-}^{(R)} = (\partial_{-}g)g^{-1}$ on
the tangent bundle of $G$. What we demonstrate is that we can take
these currents, and using the pseudoduality relations (1) we obtain
currents on $G$ (not $\tilde{G}$) and these currents are conserved.

We would like to search for infinitely many conservation
laws\cite{Pohlmeyer, Eichenherr1, Eichenherr2} on pseudodual
manifolds. We first concentrate on a simple case, where $M = G =
U(1)\times U(1)\times U(1)$ is an abelian group and $\tilde{M} =
\tilde{G}$ is $SU(2)$. We show that infinite number of conservation
laws of free scalar currents on G enable us to construct infinite
number of pseudodual current conservation on $\tilde{G}$ by means of
isometry preserving orthogonal map $T$ between tangent bundles of
these manifolds. We next focus our attention on a more complicated
case, where $M = G$ is the Lie group $SU(2)$ and $\tilde{M} =
\tilde{G}$ is $U(1)\times U(1)\times U(1)$. We find nonlocal
conserved currents on $G$ and construct pseudodual free currents on
$\tilde{G}$ using pseudoduality relations. We show that pseudodual
free scalar currents on $\tilde{G}$ gives us infinite number of
conservation laws.

\section{Pseudodual Currents : Simple Case}\label{sec:PCSC}

We take $M$ as an abelian group, and the equations of motion become
$\partial_{+-}^{2}\phi^i = 0$, where $\phi$ is free massless scalar
field. Currents on the tangent bundle of $M$ are hence given by
$J_{+}^{(L)} = (\partial_{+}\phi^{i})X_{i}$ and $J_{-}^{(R)} =
(\partial_{-}\phi^{i})X_{i}$, where ${\{X_{i}\}}$ is a basis for the
abelian Lie algebra. We notice that these currents are conserved,
$\partial_{-}J_{+}^{(L)}$ = $\partial_{+}J_{-}^{(R)}$ = $0$. Now we
take $\tilde{M}$ as a compact Lie group of the same dimensionality
with an Ad(G)-invariant metric. ${\{\tilde{X}_{i}\}}$ is the
orthonormal basis for the Lie algebra of $\tilde{G}$ with bracket
relations $[\tilde{X}_{i},\tilde{X}_{j}]_{\tilde{G}} =
\tilde{f}_{ij}^{k}\tilde{X}_{k}$, where the structure constants
$\tilde{f}_{ijk}$ are totaly antisymmetric in $ijk$. Using the map
$\tilde{g}:\Sigma\rightarrow\tilde{M}$ we may write equations of
motion as $\partial_{-}(\tilde{g}^{-1}\partial_{+}\tilde{g}) = 0$.
Currents on this manifold are defined by $\tilde{J}_{+}^{(L)} =
(\tilde{g}^{-1}\partial_{+}\tilde{g})^{i}\tilde{X}_{i}$ and
$\tilde{J}_{-}^{(R)} =
[(\partial_{-}\tilde{g})\tilde{g}^{-1}]^{i}\tilde{X}_{i}$. Again, by
virtue of equations of motion we observe that these currents are
conserved, $\partial_{-}\tilde{J}_{+}^{(L)}$ =
$\partial_{+}\tilde{J}_{-}^{(R)}$ = $0$.

To construct pseudodual currents on the manifold $M$ we make use of
the pseudoduality conditions. The pseudoduality relations between
the sigma model on an abelian group and a strict WZW sigma model on
a compact Lie group of the same dimension are

$$(\tilde{g}^{-1}\partial_{+}\tilde{g})^{i} = + T_{j}^{i}\partial_{+}\phi^{j} \eqno(2.1)
\label{pseudo1section2}$$
$$(\tilde{g}^{-1}\partial_{-}\tilde{g})^{i} =
-T_{j}^{i}\partial_{-}\phi^{j}\eqno(2.2) \label{pseudo2section2}$$
where $T$ is an orthogonal matrix and $\tilde{g}^{-1}d\tilde{g} =
(\tilde{g}^{-1}d\tilde{g})^{i}\tilde{X}_{i}$.

Taking $\partial_{-}$ of the first equation (2.1) we conclude that
$T$ is a function of $\sigma^{+}$ only. Taking $\partial_{+}$ of the
second equation (2.2) gives us the differential equation for $T$

$$[(\partial_{+}T)T^{-1}]_{j}^{i} = - \tilde{f}_{kj}^{i}T_{l}^{k}\partial_{+}\phi^{l} \eqno(3) \label{equa5}$$
where we used the antisymmetricity of $\tilde{f}_{ikj}$ at right hand side of equation.

To get pseudodual currents on the manifold $M$, we first solve this
differential equation for $T$, and then plug this into pseudoduality
equations with an initially given $\partial_{\pm}\phi^{i}$ and from
the pseudodual currents we find that these currents are conserved.

\subsection{An Example}

We consider the sigma model based on the product group $U(1)\times
U(1)\times U(1)$ for $M$ and a strict WZW model based on group
$SU(2)$ for $\tilde{M}$. We may write a point on the sigma model to
$M$ as $\phi^{i}X_{i}$, where $i=1,2,3$ and $\{X_{i}\}$ are basis.
Equations of motions are $\partial_{+-}^{2}\phi^{i} = 0$. Currents
may be written as $J_{+}^{(L)} = (\partial_{+}\phi^{i})X_{i}$ and
$J_{-}^{(R)} = (\partial_{-}\phi^{i})X_{i}$. We learn from equations
of motions that these currents are conserved.

We denote any element in $\tilde{G}$ as $\tilde{g} =
e^{i\tilde{\theta}^{k}\tilde{X}_{k}}$, where $\{\tilde{\theta}^{k}\}
= (\tilde{\theta}^{1}, \tilde{\theta}^{2},\tilde{\theta}^{3})$ and
$\{\tilde{X}_{k}\} = (-i\frac{\sigma_{1}}{2},
-i\frac{\sigma_{2}}{2}, -i\frac{\sigma_{3}}{2})$ is a basis for the
Lie algebra of $SU(2)$. Structure constants are $\epsilon_{ijk}$.
Equations of motion for the strict WZW model are
$\partial_{-}(\tilde{g}^{-1}\partial_{+}\tilde{g}) = 0$, where
$\tilde{g}^{-1}d\tilde{g} =
(\tilde{g}^{-1}d\tilde{g})^{k}\tilde{X}_{k}$. Currents for the Lie
algebra are $\tilde{J}_{+}^{(L)} =
(\tilde{g}^{-1}\partial_{+}\tilde{g})^{k}\tilde{X}_{k}$ and
$\tilde{J}_{-}^{(R)} =
[(\partial_{-}\tilde{g})\tilde{g}^{-1}]^{k}\tilde{X}_{k}$. Again
equations of motion ensure that these currents are conserved.

We first solve the ordinary differential equation for $T$ to find
the pseudodual currents. Multiplying (3) by $T_{n}^{j}$ from right
we get

$$\partial_{+}T_{n}^{i} = -\tilde{f}_{kj}^{i}T_{l}^{k}T_{n}^{j}\partial_{+}\phi^{l}\eqno(4) \label{equa6}$$
We put in an order parameter $\varepsilon$ to look for a perturbation solution,

$$\partial_{+}T_{n}^{i} = -\varepsilon\tilde{f}_{kj}^{i}T_{l}^{k}T_{n}^{j}\partial_{+}\phi^{l}\eqno(5)
\label{equa7}$$

Presumably the solution is in the form $T =
e^{\varepsilon\alpha_{1}}e^{\frac{1}{2}\varepsilon^{2}\alpha_{2}}(I
+ \mathcal{O}(\varepsilon^{3}))$, where $\alpha_{1}$ and
$\alpha_{2}$ are antisymmetric matrices. Since we know that $T$ is
only a function of $\sigma^{+}$, $\alpha_{1}$ and $\alpha_{2}$ are
also functions of $\sigma^{+}$. If we expand $T$

$$T = I + \varepsilon\alpha_{1} + \frac{1}{2}\varepsilon^{2}(\alpha_{2} + \alpha_{1}^{2}) +
\mathcal{O}(\varepsilon^{3})\eqno(6) \label{equa8}$$
then taking $\partial_{+}$ we end up with

$$\partial_{+}T = \varepsilon\partial_{+}\alpha_{1} + \frac{1}{2}\varepsilon^{2}[\alpha_{1}(\partial_{+}\alpha_{1})
+ (\partial_{+}\alpha_{1})\alpha_{1} + \partial_{+}\alpha_{2}] +
\mathcal{O}(\varepsilon^{3})\eqno(7) \label{equa9}$$ If we compare
(5) to (7), the latter may be written in tensor product form as
\begin{align}
\partial_{+}T &= -\varepsilon\tilde{f}\otimes(T\partial_{+}\phi)\otimes T = -\varepsilon\tilde{f}\otimes[(I +
\varepsilon\alpha_{1})\partial_{+}\phi]\otimes (I + \varepsilon\alpha_{1}) \notag\\ &=
-\varepsilon\tilde{f}\otimes
\partial_{+}\phi\otimes I - \varepsilon^{2}\tilde{f}\otimes \alpha_{1}\partial_{+}\phi\otimes I -
\varepsilon^{2}\tilde{f}\otimes \partial_{+}\phi\otimes\alpha_{1} \tag{8}
\end{align}
we find
$$\partial_{+}\alpha_{1} = -\tilde{f}\otimes \partial_{+}\phi\otimes I \eqno(9) \label{equa9}$$
$$\frac{1}{2}[\partial_{+}\alpha_{2} + \alpha_{1}(\partial_{+}\alpha_{1}) + (\partial_{+}\alpha_{1})\alpha_{1}] =
-\tilde{f}\otimes \alpha_{1}\partial_{+}\phi\otimes I - \tilde{f}\otimes
\partial_{+}\phi\otimes \alpha_{1} \eqno(10) \label{equa10}$$
Solving (9) we get $\alpha_{1}$ as follows

$$(\alpha_{1})_{n}^{i} = -\int_{0}^{\sigma^{+}} \tilde{f}_{kn}^{i}\partial_{+}\phi^{k} \, d\sigma'^{+} = - \tilde{f}_{kn}^{i} (\phi^{k} + C^{k}) \eqno(11)
\label{equa11}$$ where $C^{k}$ is a constant, and we choose it to be
zero. Since $\alpha_{1}$ is a function of $\sigma^{+}$ only,
$\phi^{k}$ in the expression of $\alpha_{1}$ should involve
$\sigma^{+}$, not $\sigma^{-}$. From this we understand that we need
to separate $\phi$ as right moving wave $\phi_{R}(\sigma^{-})$ and
left moving wave $\phi_{L}(\sigma^{+})$, i.e. $\phi =
\phi_{L}(\sigma^{+}) + \phi_{R}(\sigma^{-})$. Hence
$(\alpha_{1})_{n}^{i} = - \tilde{f}_{kn}^{i} \phi_{L}^{k}$, from
which we find
\[
   \alpha_{1} = \begin{pmatrix}
                   \phantom{-}0         & \phantom{-}\phi_{L}^{3}  & \phantom{-}-\phi_{L}^{2} \\
                   \phantom{-}-\phi_{L}^{3} & \phantom{-}0         &  \phantom{-}\phi_{L}^{1} \\
                              \phi_{L}^{2}  & \phantom{-}-\phi_{L}^{1} &                    0 \\
                \end{pmatrix} \eqno(12) \label{equa12}
\]
Solving (10) we obtain
\begin{align}
(\alpha_{2})_{n}^{i} &=
-\int_{0}^{\sigma^{+}}(\alpha_{1}\partial_{+}\alpha_{1})_{n}^{i} \,
d\sigma'^{+} -
\int_{0}^{\sigma^{+}}[(\partial_{+}\alpha_{1})\alpha_{1}]_{n}^{i} \,
d\sigma'^{+} \notag\\ &\ -2
\int_{0}^{\sigma^{+}}\tilde{f}_{kn}^{i}(\alpha_{1})_{l}^{k}\partial_{+}\phi_{L}^{l}
\, d\sigma'^{+}
-2\int_{0}^{\sigma^{+}}\tilde{f}_{mr}^{i}\partial_{+}\phi_{L}^{m}(\alpha_{1})_{n}^{r}
\, d\sigma'^{+} \tag{13} \label{equa13}\\ &=
-\int_{0}^{\sigma^{+}}(\phi_{L}^{i}\partial_{+}\phi_{L}^{n} -
 \phi_{L}^{n}\partial_{+}\phi_{L}^{i})\, d\sigma'^{+}\notag
\end{align}
which gives us the following entries of $\alpha_{2}$ with the help
of (12)

\begin{align}
&(\alpha_{2})_{1}^{1} = 0 \tag{14.1}\\
&(\alpha_{2})_{2}^{1} =
\int_{0}^{\sigma^{+}}[\phi_{L}^{2}(\partial_{+}\phi_{L}^{1}) -
\phi_{L}^{1}(\partial_{+}\phi_{L}^{2})] \, d\sigma'^{+} \tag{14.2}\\
&(\alpha_{2})_{3}^{1} =
\int_{0}^{\sigma^{+}}[\phi_{L}^{3}(\partial_{+}\phi_{L}^{1}) -
\phi_{L}^{1}(\partial_{+}\phi_{L}^{3})] \, d\sigma'^{+} \tag{14.3}\\
&(\alpha_{2})_{1}^{2} =
\int_{0}^{\sigma^{+}}[\phi_{L}^{1}(\partial_{+}\phi_{L}^{2}) -
\phi_{L}^{2}(\partial_{+}\phi_{L}^{1})] \, d\sigma'^{+} \tag{14.4}\\
&(\alpha_{2})_{2}^{2} = 0 \tag{14.5}\\
&(\alpha_{2})_{3}^{2} =
\int_{0}^{\sigma^{+}}[\phi_{L}^{3}(\partial_{+}\phi_{L}^{2}) -
\phi_{L}^{2}(\partial_{+}\phi_{L}^{3})] \, d\sigma'^{+} \tag{14.6}\\
&(\alpha_{2})_{1}^{3} =
\int_{0}^{\sigma^{+}}[\phi_{L}^{1}(\partial_{+}\phi_{L}^{3}) -
\phi_{L}^{3}(\partial_{+}\phi_{L}^{1})] \, d\sigma'^{+} \tag{14.7}\\
&(\alpha_{2})_{2}^{3} =
\int_{0}^{\sigma^{+}}[\phi_{L}^{2}(\partial_{+}\phi_{L}^{3}) -
\phi_{L}^{3}(\partial_{+}\phi_{L}^{2})] \, d\sigma'^{+} \tag{14.8}\\
&(\alpha_{2})_{3}^{3} = 0 \tag{14.9} \label{equa14}
\end{align}

Plugging (12) and (14) into $T$ and setting $\varepsilon = 1$ gives
us

\begin{align}
T_{j}^{i} &= \delta_{j}^{i} + (\alpha_{1})_{j}^{i} +
\frac{1}{2}[(\alpha_{2})_{j}^{i} +
(\alpha_{1}^{2})_{j}^{i}] + \mathcal{O}(\phi^{3}) \tag{15}\\
&\approx\delta_{j}^{i} - \tilde{f}_{kj}^{i} \phi_{L}^{k} -
\frac{1}{2}[\int_{0}^{\sigma^{+}}(\phi_{L}^{i}\partial_{+}\phi_{L}^{j}
- \phi_{L}^{j}\partial_{+}\phi_{L}^{i})\, d\sigma'^{+} -
\tilde{f}_{km}^{i} \tilde{f}_{nj}^{m}\phi_{L}^{k}\phi_{L}^{n}]
\notag
\end{align}
so the entries of $T$ becomes
\begin{align}
&T_{1}^{1} = 1 - \frac{1}{2}[\phi_{L}^{2}\phi_{L}^{2} + \phi_{L}^{3}\phi_{L}^{3}] \tag{16.1}\\
&T_{2}^{1} = \phi_{L}^{3} + \int_{0}^{\sigma^{+}}\phi_{L}^{2}(\partial_{+}\phi_{L}^{1}) \, d\sigma'^{+} \tag{16.2}\\
&T_{3}^{1} = -\phi_{L}^{2} +
\int_{0}^{\sigma^{+}}\phi_{L}^{3}(\partial_{+}\phi_{L}^{1}) \,
d\sigma'^{+} \tag{16.3}\\
&T_{1}^{2} = -\phi_{L}^{3} +
\int_{0}^{\sigma^{+}}\phi_{L}^{1}(\partial_{+}\phi_{L}^{2}) \,
d\sigma'^{+} \tag{16.4}
\end{align}
\begin{align}
&T_{2}^{2} = 1 - \frac{1}{2}[\phi_{L}^{1}\phi_{L}^{1} + \phi_{L}^{3}\phi_{L}^{3}] \tag{16.5}\\
&T_{3}^{2} = \phi_{L}^{1} +
\int_{0}^{\sigma^{+}}\phi_{L}^{3}(\partial_{+}\phi_{L}^{2}) \,
d\sigma'^{+} \tag{16.6}\\
&T_{1}^{3} = \phi_{L}^{2} + \int_{0}^{\sigma^{+}}\phi_{L}^{1}(\partial_{+}\phi_{L}^{3}) \, d\sigma'^{+} \tag{16.7}\\
&T_{2}^{3} = -\phi_{L}^{1} + \int_{0}^{\sigma^{+}}\phi_{L}^{2}(\partial_{+}\phi_{L}^{3}) \, d\sigma'^{+} \tag{16.8}\\
&T_{3}^{3} = 1 - \frac{1}{2}[\phi_{L}^{1}\phi_{L}^{1} +
\phi_{L}^{2}\phi_{L}^{2}] \tag{16.9} \label{equa16}
\end{align}

We note that $T$ is an orthogonal matrix. The type of the field
$\phi(\sigma^{+}, \sigma^{-}) = \phi_{L}(\sigma^{+}) +
\phi_{R}(\sigma^{-})$ puts pseudoduality relations into the forms
\begin{align}
&(\tilde{g}^{-1}\partial_{+}\tilde{g})^{i} = +T_{j}^{i}\partial_{+}\phi_{L}^{j} \tag{17.1}\label{pseudo2section3}\\
&(\tilde{g}^{-1}\partial_{-}\tilde{g})^{i} =
-T_{j}^{i}\partial_{-}\phi_{R}^{j} \tag{17.2}
\label{pseudo2section4}
\end{align}
We note that equation (17.1) has an invariance under
$\tilde{g}(\sigma^{+}, \sigma^{-})$ $\longrightarrow$
$h(\sigma^{-})\tilde{g}(\sigma^{+}, \sigma^{-})$. From this we can
look for solution $\tilde{g}(\sigma^{+}, \sigma^{-}) =
\tilde{g}_{R}(\sigma^{-})\tilde{g}_{L}(\sigma^{+})$, so first
pseudoduality relation is reduced to
$(\tilde{g}_{L}^{-1}\partial_{+}\tilde{g}_{L})^{i} =
+T_{j}^{i}\partial_{+}\phi_{L}^{j}$. This equation gives us the left
current. Next we have to find $\tilde{g}_{R}(\sigma^{-})$ using
second pseudoduality equation to construct right current. Plugging
$\tilde{g}(\sigma^{+}, \sigma^{-}) =
\tilde{g}_{R}(\sigma^{-})\tilde{g}_{L}(\sigma^{+})$ into (17.2) and
arranging terms we obtain
$$\tilde{g}_{R}^{-1}(\sigma^{-})\partial_{-}\tilde{g}_{R}(\sigma^{-})
=
-\tilde{g}_{L}(\sigma^{+})(\tilde{X}_{i}T_{j}^{i}\partial_{-}\phi_{R}^{j})\tilde{g}_{L}^{-1}(\sigma^{+})\eqno(18)
\label{equa18}$$ where $\{\tilde{X}_{i}\}$ are the Lie algebra basis
of $\mathbf{\tilde{g}}$, and $(\tilde{X}_{i})_{jk} =
\epsilon_{jik}$. Since we want to construct  pseudodual currents in
the order of $\phi^{n}$, we need $T(\sigma^{+})$ to the order of
$\phi^{n-1}$ to get $\tilde{J}_{+}^{(L)}(\sigma^{+})$ to the order
of $\phi^{n}$. From equation (18) we see that the knowledge of $T$
to $\mathcal{O}(\phi^{n-1})$ and $\tilde{g}_{L}$ to
$\mathcal{O}(\phi^{n-1})$ allows us to construct $\tilde{g}_{R}$ to
$\mathcal{O}(\phi^{n})$, so we can construct
$\tilde{J}_{-}^{(R)}(\sigma^{-})$ to $\mathcal{O}(\phi^{n})$.

First we construct $\tilde{J}_{+}^{(L)}(\sigma^{+})$ to the order of
$\phi^{2}$, so we need $T$ to $\mathcal{O}(\phi)$

$$T_{j}^{i} = \delta_{j}^{i} - \tilde{f}_{kj}^{i}
\phi_{L}^{k} + \mathcal{O}(\phi^{2}) \eqno(19)$$ Therefore, using
first pseudoduality relation (17.1)
$$\tilde{J}_{+}^{(L)}(\sigma^{+}) = \tilde{g}_{L}^{-1}\partial_{+}\tilde{g}_{L}
= \tilde{X}_{i}\partial_{+}\phi_{L}^{i} -
\tilde{X}_{i}\tilde{f}_{kj}^{i}
\phi_{L}^{k}\partial_{+}\phi_{L}^{j}\eqno(20)\label{equation20}\\$$
All we need is $\tilde{g}_{L}$ to the order of $\phi$, so we need to
solve $\tilde{g}_{L}^{-1}\partial_{+}\tilde{g}_{L} =
\tilde{X}_{i}\partial_{+}\phi_{L}^{i}$ for
$\tilde{g}_{L}(\sigma^{+})$. Choosing initial condition as
$\tilde{g}_{L}(\sigma^{+} = 0) = I$, we get
$$\tilde{g}_{L}(\sigma^{+}) = I + \tilde{X}_{i}\phi_{L}^{i} +
\mathcal{O}(\phi^{2}) \eqno(21)\\$$ Its inverse is
$$\tilde{g}_{L}^{-1}(\sigma^{+}) = I - \tilde{X}_{i}\phi_{L}^{i} +
\mathcal{O}(\phi^{2}) \eqno(22)\\$$ Plugging these into (18) we find
\begin{align}
\tilde{g}_{R}^{-1}\partial_{-}\tilde{g}_{R} &= -(I +
\tilde{X}_{l}\phi_{L}^{l})\tilde{X}_{i}(\delta_{j}^{i} -
\tilde{f}_{kj}^{i} \phi_{L}^{k})\partial_{-}\phi_{R}^{j}(I -
\tilde{X}_{k}\phi_{L}^{k})\notag\\ &=
-\tilde{X}_{i}\partial_{-}\phi_{R}^{i} +
\mathcal{O}(\phi^{3})\tag{23}\label{equation23}
\end{align}
We notice that the order of $\phi^{2}$ terms are cancelled, and
$\tilde{g}_{R}$ is a function of $\sigma^{-}$ only. We let
$\tilde{g}_{R} =
e^{-\phi_{R}^{i}\tilde{X}_{i}}e^{\xi^{k}\tilde{X}_{k}}$, where $\xi$
represents $\mathcal{O}(\phi^{2})$. Expanding $\tilde{g}_{R}$
\begin{align}
\tilde{g}_{R} &= (I - \phi_{R}^{i}\tilde{X}_{i} +
\frac{1}{2}\phi_{R}^{i}\phi_{R}^{j}\tilde{X}_{i}\tilde{X}_{j})(I +
\xi^{k}\tilde{X}_{k}) \notag \\ &= I - \phi_{R}^{i}\tilde{X}_{i} +
\frac{1}{2}\phi_{R}^{i}\phi_{R}^{j}\tilde{X}_{i}\tilde{X}_{j} +
\xi^{k}\tilde{X}_{k} +
\mathcal{O}(\phi^{3})\tag{24}\label{equation24}
\end{align}
the inverse $\tilde{g}_{R}^{-1}$ can be found from $\tilde{g}_{R} =
e^{-\xi^{k}\tilde{X}_{k}}e^{\phi_{R}^{i}\tilde{X}_{i}}$
\begin{align}
\tilde{g}_{R}^{-1} &= (I - \xi^{k}\tilde{X}_{k})(I +
\phi_{R}^{i}\tilde{X}_{i} +
\frac{1}{2}\phi_{R}^{i}\phi_{R}^{j}\tilde{X}_{i}\tilde{X}_{j})\notag\\
& = I + \phi_{R}^{i}\tilde{X}_{i} +
\frac{1}{2}\phi_{R}^{i}\phi_{R}^{j}\tilde{X}_{i}\tilde{X}_{j} -
\xi^{k}\tilde{X}_{k} + \mathcal{O}(\phi^{3})
\tag{25}\label{equation25}
\end{align}
It follows then that equations (24) and (25) lead to
$$\tilde{g}_{R}^{-1}\partial_{-}\tilde{g}_{R} =
-\partial_{-}\phi_{R}^{i}\tilde{X}_{i} +
\partial_{-}\xi^{k}\tilde{X}_{k} + \mathcal{O}(\phi^{3})\eqno(26)\label{equation26}$$
and comparison with (23) evaluates $\partial_{-}\xi^{k} = 0$, so
$\xi^{k}$ is constant and we choose it to be zero. Therefore, right
current can be constructed using (24) and (25) as
$$\tilde{J}_{-}^{(R)}(\sigma^{-}) = (\partial_{-}\tilde{g}_{R})\tilde{g}_{R}^{-1} =
-(\partial_{-}\phi_{R}^{i})\tilde{X}_{i}\eqno(27)\label{equation27}$$
we see that order of $\phi^{2}$ disappears in the expression of
right current. If we explicitly write pseudodual currents on the
manifold $M$ up to the order of $\phi^{3}$ using equations (20) and
(27) we get the following
\begin{align}
\tilde{J}_{+}^{(L)}(\sigma^{+}) =
&\tilde{X}_{i}[\partial_{+}\phi_{L}^{i} -
\tilde{f}_{kj}^{i}\phi_{L}^{k}\partial_{+}\phi_{L}^{j} -
\frac{1}{2}[\int_{0}^{\sigma^{+}}(\phi_{L}^{i}\partial_{+}\phi_{L}^{j}
- \phi_{L}^{j}\partial_{+}\phi_{L}^{i})\, d\sigma'^{+} \notag\\&-
\tilde{f}_{km}^{i}\tilde{f}_{nj}^{m}\phi_{L}^{k}\phi_{L}^{n}]\partial_{+}\phi_{L}^{j}]\tag{28.1}
\end{align}
\begin{equation}
\tilde{J}_{-}^{(R)}(\sigma^{-}) = -
\tilde{X}_{i}(\partial_{-}\phi_{R}^{i})\tag{28.2}
\end{equation}
Therefore, our currents can be written as
\begin{equation}
\tilde{J}^{(\mu)} = \tilde{J}_{[0]}^{(\mu)} +
\tilde{J}_{[1]}^{(\mu)} + \tilde{J}_{[2]}^{(\mu)} +
\mathcal{O}(\phi^{3}) \tag{29}
\end{equation}
where $\{\mu\} = (R, L)$. We can organize all these terms as
\begin{equation}
\tilde{J}^{(\mu)}(\phi) =
\sum_{0}^{\infty}\tilde{J}_{[n]}^{(\mu)}(\phi) \tag{30}
\end{equation}

It is easy to see that these currents are conserved, i.e.
$\partial_{+}\tilde{J}_{-}^{(R)}$ =
$\partial_{-}\tilde{J}_{+}^{(L)}$ = $0$, by means of the equations
of motion $\partial_{+-}^{2}\phi^{i} = 0$. Since each term satisfies
$\partial_{+}\tilde{J}_{[n]}^{(R)}$ =
$\partial_{-}\tilde{J}_{[n]}^{(L)}$ = $0$ for all $n$ separately, we
have infinite number of conservation laws for each order of $\phi$
as pointed out in~\cite{Alvarez1}.

\section{Pseudodual Currents : Complicated Case}\label{sec:PCCC}

In this case we consider the pseudoduality between two strict WZW
models based on compact Lie groups of dimension $n$ with
Ad-invariant metrics. If $\{X_{i}\}$ are the orthonormal basis for
the Lie algebra of $G$ with commutation relations $[X_{i},X_{j}]_{G}
= f_{ij}^{k}X_{k}$, where $f_{ijk}$ are totally antisymmetric in
$ijk$, and $g : \Sigma \rightarrow M$ is the map to the target
space, we may write equations of motion on $G$ as
$\partial_{-}(g^{-1}\partial_{+}g) = 0$. Therefore, currents become
$J_{+}^{(L)} = (g^{-1}\partial_{+}g)^{i}X_{i}$ and $J_{-}^{(R)} =
[(\partial_{-}g)g^{-1}]^{i}X_{i}$. These currents are conserved. We
make similar assumptions for the Lie group $\tilde{G}$. The
pseudoduality equations are
\begin{equation}
(\tilde{g}^{-1}\partial_{+}\tilde{g})^{i} =
+T_{j}^{i}(g^{-1}\partial_{+}g)^{j} \tag{31.1} \label{equa31.1}
\end{equation}
\begin{equation}
(\tilde{g}^{-1}\partial_{-}\tilde{g})^{i} =
-T_{j}^{i}(g^{-1}\partial_{-}g)^{j} \tag{31.2} \label{equa31.2}
\end{equation}
where $T$ is an orthogonal matrix. Taking $\partial_{-}$ of the
first equation (31.1) we learn that $T$ is a function of
$\sigma^{+}$ only. Taking $\partial_{+}$ of the second equation
(31.2) we get the differential equation for $T$

\begin{equation}
[(\partial_{+}T)T^{-1}]_{j}^{i} =
-\tilde{f}_{kj}^{i}T_{k}^{l}(g^{-1}\partial_{+}g)^{l} +
f_{ml}^{k}T_{k}^{i}T_{l}^{j}(g^{-1}\partial_{+}g)^{m} \tag{32}
\label{equa32}
\end{equation}

We follow the same method as we did in the previous part to find
pseudodual currents. We first solve differential equation (32) for
$T$, then replace this into the pseudoduality relations, and finally
build pseudodual currents. We will see that these currents are
conserved.

\subsection{An Example}

To illustrate all these steps in an example, we consider a strict
WZW model based on Lie group $SU(2)$ for $G$, and a sigma model
based on abelian group $U(1)\times U(1)\times U(1)$ for $\tilde{G}$.
Using the map $g : \Sigma \rightarrow G$, we may represent any
element in $G$ by $g = e^{i\phi^{k}X_{k}}$, where $\{\phi^{k}\} =
(\phi^{1}, \phi^{2}, \phi^{3})$ are commuting fields and $\{X_{k}\}$
are the orthonormal basis for the Lie algebra of $G$, and $\{X_{k}\}
= (-\frac{i}{2}\sigma_{1}, -\frac{i}{2}\sigma_{2},
-\frac{i}{2}\sigma_{3})$ for the case of $SU(2)$. Structure
constants are $\epsilon_{jk}^{i}$, and commutation relations are the
familiar form of Pauli matrices,
$[-i\frac{\sigma_{i}}{2},-i\frac{\sigma_{j}}{2}] =
\epsilon_{ij}^{k}(-i\frac{\sigma_{k}}{2})$. Equations of motion are
$\partial_{-}(g^{-1}\partial_{+}g) = 0$. Nonlocal currents for the
Lie algebra of $SU(2)$ are $J_{+}^{(L)} =
(g^{-1}\partial_{+}g)^{k}X_{k}$ and $J_{-}^{(R)} =
[(\partial_{-}g)g^{-1}]^{k}X_{k}$. We want to construct currents up
to the order of $\phi^{2}$. If we consider infinitesimal
coefficients $\{\phi^{k}\}$, keeping up to second orders we may
expand $g$ as

\begin{equation}
g = 1 + i\phi^{k}X_{k} - \frac{1}{2}(\phi^{k}\phi^{l})(X_{k}X_{l}) +
... \tag{33}
\end{equation}
Since we are looking for $J_{+}^{(L)}$ and $J_{-}^{(R)}$ up to the
order of $\phi^{2}$, we need $g$ to the order of $\phi$, hence
\begin{align}
g &= 1 + i\phi^{k}X_{k}\tag{34.1}\label{34.1}\\
g^{-1} &= 1 - i\phi^{k}X_{k}\tag{34.2}\label{34.2}
\end{align}
To this order the solution to equations of motion
$\partial_{-}(g^{-1}\partial_{+}g) = 0$ is $g =
g_{R}(\sigma^{-})g_{L}(\sigma^{+})$, which leads to
$\phi(\sigma^{+}, \sigma^{-}) = \phi_{R}(\sigma^{-}) +
\phi_{L}(\sigma^{+})$. Thus equation (34) can be written as
$$g_{L} = 1 + i\phi_{L}^{k}X_{k}\eqno(35.1)\label{35.1}$$
$$g_{R} = 1 + i\phi_{R}^{k}X_{k}\eqno(35.2)\label{35.2}$$
 and hence left and right currents can readily be obtained as
 \begin{align}
&J_{+}^{(L)} = g_{L}^{-1}\partial_{+}g_{L} =
i\partial_{+}\phi_{L}^{m}X_{m} +
\frac{1}{2}f_{kl}^{m}\phi_{L}^{k}\partial_{+}\phi_{L}^{l}X_{m}\tag{36.2}\\
 &J_{-}^{(R)} = (\partial_{-}g_{R})g_{R}^{-1} = i\partial_{-}\phi_{R}^{m}X_{m} +
 \frac{1}{2}f_{kl}^{m}\phi_{R}^{l}\partial_{-}\phi_{R}^{k}X_{m}\tag{36.2}
\end{align}

Therefore, we conclude that $\partial_{-}J_{+}^{(L)}$ =
$\partial_{+}J_{-}^{(R)}$ = $0$, i.e, currents are conserved on $G$.
We first solve equation (32) to figure out the pseudodual currents.
Since $\tilde{f}_{kj}^{i} = 0$, we have

\begin{equation}
[(\partial_{+}T)T^{-1}]_{j}^{i} =
f_{ml}^{k}T_{k}^{i}T_{l}^{j}(g_{L}^{-1}\partial_{+}g_{L})^{m}
\tag{37}
\end{equation}
this may be reduced to

\begin{equation}
[(\partial_{+}T)]_{n}^{i} =
f_{mn}^{k}T_{k}^{i}(g_{L}^{-1}\partial_{+}g_{L})^{m} \tag{38}
\end{equation}
and putting in an order parameter $\varepsilon$ we get
\begin{equation}
[(\partial_{+}T)]_{n}^{i} = \varepsilon
f_{mn}^{k}T_{k}^{i}(g_{L}^{-1}\partial_{+}g_{L})^{m} \tag{39}
\label{equa39}
\end{equation}

We adapt to an exponential solution $T = e^{\varepsilon
\alpha_{1}}e^{\frac{1}{2}\varepsilon^{2}\alpha_{2}}(I +
\mathcal{O}(\varepsilon^{3}))$, where $\alpha_{1}$ and $\alpha_{2}$
are antisymmmetric matrices, and expanding this solution we get

$$T = I + \varepsilon\alpha_{1} + \frac{1}{2}\varepsilon^{2}(\alpha_{2} + \alpha_{1}^{2}) +
\mathcal{O}(\varepsilon^{3})\eqno(40) \label{equa40}$$ taking
$\partial_{+}$ of $T$ leads to
$$\partial_{+}T = \varepsilon\partial_{+}\alpha_{1} + \frac{1}{2}\varepsilon^{2}[\alpha_{1}(\partial_{+}\alpha_{1})
+ (\partial_{+}\alpha_{1})\alpha_{1} + \partial_{+}\alpha_{2}] +
\mathcal{O}(\varepsilon^{3})\eqno(41) \label{equa41}$$ expressing
(39) in tensor product form
\begin{align}
\partial_{+}T &= \varepsilon f(g_{L}^{-1}\partial_{+}g_{L})\otimes T = \varepsilon f(g_{L}^{-1}\partial_{+}g_{L})\otimes (I + \varepsilon\alpha_{1})  \notag\\ &= \varepsilon f(g_{L}^{-1}\partial_{+}g_{L}) + \varepsilon^{2}f(g_{L}^{-1}\partial_{+}g_{L})\otimes \alpha_{1}\tag{42}
\end{align}
and comparing this with (41) we obtain $\alpha_{1}$
\begin{align}
(\alpha_{1})_{n}^{i} &=
\int_{0}^{\sigma^{+}}f_{mn}^{i}(g_{L}^{-1}\partial_{+}g_{L})^{m}d\sigma'^{+}
\tag{43}\label{equa43}\\ &=if_{mn}^{i}\phi_{L}^{m} +
\frac{1}{2}f_{mn}^{i}f_{kl}^{m}\int_{0}^{\sigma^{+}}\phi_{L}^{k}\partial_{+}\phi_{L}^{l}\,d\sigma'^{+}\notag\\
&= i\epsilon_{mn}^{i}\phi_{L}^{m} +
\frac{1}{2}\int_{0}^{\sigma^{+}}(\phi_{L}^{n}\partial_{+}\phi_{L}^{i}
- \phi_{L}^{i}\partial_{+}\phi_{L}^{n})\, d\sigma'^{+}\notag
\end{align}
this expression leads to the following entries

\begin{align}
&(\alpha_{1})_{1}^{1} = 0 \tag{44.1}\\ &(\alpha_{1})_{2}^{1} =
-i\phi_{L}^{3}-\frac{1}{2}\int_{0}^{\sigma^{+}}[\phi_{L}^{1}(\partial_{+}\phi_{L}^{2})
- \phi_{L}^{2}(\partial_{+}\phi_{L}^{1})] \, d\sigma'^{+} \tag{44.2}\\
&(\alpha_{1})_{3}^{1} =
i\phi_{L}^{2}+\frac{1}{2}\int_{0}^{\sigma^{+}}[\phi_{L}^{3}(\partial_{+}\phi_{L}^{1})
- \phi_{L}^{1}(\partial_{+}\phi_{L}^{3})] \, d\sigma'^{+} \tag{44.3}
\end{align}
\begin{align}
&(\alpha_{1})_{1}^{2} =
i\phi_{L}^{3}+\frac{1}{2}\int_{0}^{\sigma^{+}}[\phi_{L}^{1}(\partial_{+}\phi_{L}^{2})
- \phi_{L}^{2}(\partial_{+}\phi_{L}^{1})] \, d\sigma'^{+}
\tag{44.4}\\
&(\alpha_{1})_{2}^{2} = 0 \tag{44.5}\\
&(\alpha_{1})_{3}^{2} =
-i\phi_{L}^{1}-\frac{1}{2}\int_{0}^{\sigma^{+}}[\phi_{L}^{2}(\partial_{+}\phi_{L}^{3})
- \phi_{L}^{3}(\partial_{+}\phi_{L}^{2})] \, d\sigma'^{+} \tag{44.6}\\
&(\alpha_{1})_{1}^{3} =
-i\phi_{L}^{2}-\frac{1}{2}\int_{0}^{\sigma^{+}}[\phi_{L}^{3}(\partial_{+}\phi_{L}^{1})
- \phi_{L}^{1}(\partial_{+}\phi_{L}^{3})] \, d\sigma'^{+} \tag{44.7}\\
&(\alpha_{1})_{2}^{3} =
i\phi_{L}^{1}+\frac{1}{2}\int_{0}^{\sigma^{+}}[\phi_{L}^{2}(\partial_{+}\phi_{L}^{3})
- \phi_{L}^{3}(\partial_{+}\phi_{L}^{2})] \, d\sigma'^{+} \tag{44.8}\\
&(\alpha_{1})_{3}^{3} = 0 \tag{44.9}
\end{align}
and
$$\partial_{+}\alpha_{2} = 2f(g_{L}^{-1}\partial_{+}g_{L})\otimes
\alpha_{1} - [\alpha_{1}(\partial_{+}\alpha_{1}) +
(\partial_{+}\alpha_{1})\alpha_{1}]$$ Hence, $\alpha_{2}$ is
obtained as
$$(\alpha_{2})_{n}^{i} = \int_{0}^{\sigma^{+}} (\phi_{L}^{i}\partial_{+}\phi_{L}^{n} -
\phi_{L}^{n}\partial_{+}\phi_{L}^{i})\, d\sigma'^{+}\eqno(45)
\label{equa45}$$ We see that this is equivalent to (13), and entries
are the same as the negative of (14). Therefore, we can find $T$ by
means of (43) and (45), and setting $\varepsilon = 1$

\begin{align}
T_{n}^{i} = \delta_{n}^{i} + i\epsilon_{mn}^{i}\phi_{L}^{m} +
\frac{1}{2}(\delta_{n}^{i}\phi_{L}^{m}\phi_{L}^{m} -
\phi_{L}^{n}\phi_{L}^{i}) \tag{46}
\end{align}

Again we note that $T$ is an orthogonal matrix. Now using
pseudoduality equations (31)

\begin{equation}
\partial_{+}\tilde{\phi_{L}}^{i} =
+T_{j}^{i}(g^{-1}\partial_{+}g)^{j} =
+T_{j}^{i}(g_{L}^{-1}\partial_{+}g_{L})^{j} \tag{47.1}
\end{equation}

\begin{equation}
\partial_{-}\tilde{\phi_{R}}^{i} = -T_{j}^{i}(g^{-1}\partial_{-}g)^{j} = -T_{j}^{i}(g_{L}^{-1}g_{R}^{-1}(\partial_{-}g_{R})g_{L})^{j}  \tag{47.2}
\end{equation}
since we are trying to find $\partial_{+}\tilde{\phi_{L}}^{i}$ and
$\partial_{-}\tilde{\phi_{R}}^{i}$ up to the order of $\phi^{2}$, we
need $T$ to the order of $\phi$, hence using
$$T_{n}^{i} = \delta_{j}^{i} + i\epsilon_{mj}^{i}\phi_{L}^{m} +
\mathcal{O}(\phi^{2})\eqno(48)$$
\begin{align}
\partial_{+}\tilde{\phi_{L}}^{i} &=
T_{j}^{i}(g_{L}^{-1}\partial_{+}g_{L})^{j}\notag\\ &=
[\delta_{j}^{i} +
i\epsilon_{mj}^{i}\phi_{L}^{m}][i\partial_{+}\phi_{L}^{j} +
\frac{1}{2}\epsilon_{kl}^{j}\phi_{L}^{k}\partial_{+}\phi_{L}^{l}]\notag\\
&= i\partial_{+}\phi_{L}^{i} -
\frac{1}{2}\epsilon_{kl}^{i}\phi_{L}^{k}\partial_{+}\phi_{L}^{l} +
\mathcal{O}(\phi^{3})\tag{49.1}
\end{align}

\begin{align}
\partial_{-}\tilde{\phi_{R}}^{i} &= -
T_{j}^{i}(g^{-1}\partial_{-}g)^{j} = -
T_{j}^{i}[g_{L}^{-1}(g_{R}^{-1}\partial_{-}g_{R})g_{L}]^{j}\notag\\
&= -[\delta_{j}^{i} + i\epsilon_{mj}^{i}\phi_{L}^{m}][(1-i\phi_{L}^{k}X_{k})(1-\phi_{R}^{k}X_{k})(i\partial_{-}\phi_{R}^{k}X_{k} - \frac{1}{2}\phi_{R}^{m}\partial_{-}\phi_{R}^{l}X_{l}X_{m} \notag\\ &- \frac{1}{2}\phi_{R}^{l}\partial_{-}\phi_{R}^{m}X_{l}X_{m})(1 + i\phi_{L}^{k}X_{k})]^{j} \notag\\
&= -[\delta_{j}^{i} +
i\epsilon_{mj}^{i}\phi_{L}^{m}][i\partial_{-}\phi_{R}^{j} +
 \epsilon_{mn}^{j}\phi_{L}^{m}\partial_{-}\phi_{R}^{n} + \frac{1}{2}\epsilon_{lm}^{j}\phi_{R}^{l}\partial_{-}\phi_{R}^{m}]\notag\\
 &= -i\partial_{-}\phi_{R}^{i} - \frac{1}{2}\epsilon_{lm}^{i}\phi_{R}^{l}\partial_{-}\phi_{R}^{m}\tag{49.2}
\end{align}

We see that these are free scalar currents on the tangent bundle to
pseudodual manifold $\tilde{G}$. Since
$\partial_{+}\tilde{\phi_{L}}^{i}$ depends only on $\sigma^{+}$, and
$\partial_{-}\tilde{\phi_{R}}^{i}$ only on $\sigma^{-}$, these
pseudodual free scalar currents are conserved provided that
equations of motion for free scalar fields hold. We go back to
equations of motion to see that these pseudodual tangent bundle
components take us to pseudodual conserved currents. Equations of
motion, $\partial_{-}(g^{-1}\partial_{+}g)^{i} = 0$, imply that
$\partial_{+-}^{2}\phi^{i} = 0$. Obviously we find out the
pseudodual conservation laws $\partial_{\pm\mp}^{2}\tilde{\phi}^{i}
= 0$ in all $\phi$-orders using these conditions.

\section{Discussion}\label{sec:int}

We observed that nonlinear character of WZW models results in an
infinite number of terms in transformation matrix $T$, which in turn
leads to construct infinite number of nonlocal currents in
pseudodual manifold. Calculations here were motivated by the fact
that sigma models have Lie group structures, and $T \in SO(n)$.
Hence structure of Lie groups together with perturbation
calculations reflects the nonlinear characteristic of sigma models.
It is obvious that pseudoduality transformation preserved the
pseudodual currents in our cases where one model based on an abelian
group $U(1)\times U(1)\times U(1)$ in two cases we discussed.
However, One can consider general Lie group valued fields for both
models, and see that this would also yield conserved currents on
pseudodual model. We considered three dimensional models for
simplicity but this can be extended to any dimension. Calculation of
these currents gives us curvatures by means of Cartan structural
equations
\begin{align}
dw^{i} + w_{j}^{i} \wedge w^{j} &= 0 \tag{50.1}\\
dw_{j}^{i} + w_{k}^{i} \wedge w_{j}^{k} &=
\frac{1}{2}R_{jkl}^{i}w^{k} \wedge w^{l} \tag{50.2}
\end{align}
where $w^{i} = J^{i}$ and $w_{j}^{i} = \frac{1}{2}f_{kj}^{i}J^{k}$
is the antisymmetric connection, and $J$ stands for both
$J_{+}^{(L)}$ and $J_{-}^{(R)}$. These currents form an orthonormal
frame on pullback bundle $g^{*}(TG)$. Since we considered abelian
models, and hence obtained scalar currents, it is easily noted that
curvatures are zero. In general case where sigma models based on
general Lie groups, curvatures will be constant and opposite. This
shows that sigma models are based on symmetric spaces as pointed out
in \cite{Alvarez2}. The calculations and results of this paper can
be applied to pseudoduality relations between symmetric space sigma
models to construct nonlocal currents, and as a result curvatures.
In more extreme cases it would be interesting to discuss the
pseudoduality in supersymmetric sigma models, and get constraints
for pseudoduality transformation.

\section*{Acknowledgments}

I would like to thank O. Alvarez for his comments, helpful
discussions, and reading an earlier draft of the manuscript.


\begin{thebibliography}{9}

\bibitem{Alvarez1}
  O. Alvarez,
  \emph{Pseudoduality in Sigma Models},
  Nucl.Phys. B638 (2002) 328-350, \href{http://www.arxiv.org/pdf/hep-th/0204011/}{hep-th/0204011/}.

\bibitem{Curtright1}
  T. Curtright and C. Zachos,
  \emph{Currents, charges, and canonical structure of pseudodual chiral
  models},
  Phys. Rev. D49 (1994) 5408-5421,
  \href{http://www.arxiv.org/pdf/hep-th/9401006/}{hep-th/9401006/}.


\bibitem{Alvarez2}
  O. Alvarez,
  \emph{Target space pseudoduality between dual symmetric spaces},
  Nucl. Phys. B582 (2000) 139, \href{http://www.arxiv.org/pdf/hep-th/0004120/}{hep-th/0004120/}.

\bibitem{Witten}
  E. Witten,
  \emph{Nonabelian bozonization in two dimensions},
  Commun. Math. Phys. 92 (1984) 455-472.

\bibitem{Evans1}
  J. M. Evans, M. Hassan, N. J. MacKay, and A. J. Mountain,
  \emph{Conserved charges and supersymmetry in principal chiral and WZW models},
  Nucl.Phys. B580 (2000) 605-646, \href{http://arxiv.org/abs/hep-th/0001222/}{hep-th/0001222/}.

\bibitem{Evans2}
  J. M. Evans, M. Hassan, N. J. MacKay, and A. J. Mountain,
  \emph{Local conserved charges in principal chiral models},
  Nucl. Phys. B561 (1999) 385-412,
  \href{http://www.arxiv.org/pdf/hep-th/9902008/}{hep-th/9902008/}.

\bibitem{Ivanov}
  E. A. Ivanov,
  \emph{Duality in d = 2 sigma models of chiral field with anomaly},
  Theor. Math. Phys. 71 (1987 474-484).

\bibitem{Pohlmeyer}
  K. Pohlmeyer,
  \emph{Integrable hamiltonian systems and interactions through quadratic
  constraints},
  Commun. Math. Phys. 46 (1976) 207.

\bibitem{Eichenherr1}
  H. Eichenherr and M. Forger,
  \emph{On the dual symmetry of nonlinear sigma models},
  Nucl. Phys. B155 (1979) 381.

\bibitem{Eichenherr2}
  H. Eichenherr and M. Forger,
  \emph{Higher local conservation laws for nonlinear sigma models on symmetric
  spaces},
  Commun. Math. Phys. 82 (1981) 227.





\end{thebibliography}
\end{document}